\documentclass[preprint,superscriptaddress,amsmath,amssymb,pra]{revtex4-1}
\usepackage{graphicx,dcolumn,bm}

\begin{document}	
	\title{Sensing single atoms in a cavity using a broadband squeezed light}
	\author{D. Q. Bao}
	\author{C. J. Zhu}
	\email{cjzhu@tongji.edu.cn}
	\author{Y. P. Yang}
	\email{yang\_yaping@tongji.edu.cn}
	\affiliation{MOE Key Laboratory of Advanced Micro-Structured Materials, School of Physics Science and Engineering, Tongji University, Shanghai, China 200092.}
	\author{G. S. Agarwal}
	\email{girish.agarwal@tamu.edu}
	\affiliation{Institute for QuantumScience and Engineering, and Department of Biological and Agricultural Engineering Texas A\&M University, College Station, Texas 77843, USA}
	\date{\today}
	
	\begin{abstract}
	We investigate a single atom cavity-QED system directly driven by a broadband squeezed light. We demonstrate how the squeezed radiation can be used to sense the presence of a single atom in a cavity. This happens by transferring one of the photons from the field in a state with even number of photons to the atom and thereby populating odd number Fock states. Specifically, the presence of the atom is sensed by remarkable changing in the presence of one photon and the loss of squeezing of the cavity field. A complete study of quantum fluctuations and the excitation of multiphoton transitions is given.
	\end{abstract}
\maketitle
\section{Introduction}
Cavity quantum electrodynamics (cavity QED) is the study on the interaction between photons confined to a high-finesse cavity and quantum emitters, including atoms, quantum dots and so on, under conditions where the quantum nature of cavity photons is dominant~\cite{dutra2005cavity}. The typical model of light-matter interaction is a single atom interacting with photons in the cavity, which is known as the Jaynes-Cummings model and provides many interesting quantum optical phenomena, including vacuum Rabi splitting~\cite{yoshie2004vacuum}, single photon blockade~\cite{birnbaum2005photon} and so on~\cite{scully1997quantum,agarwal2012quantum}. It also provides possibilities to achieve quantum information processing and communication by fully controlling the cavity QED systems~\cite{imamog1999quantum,kimble2008quantum}.
Due to the quantum nature and extremely high coherence, atoms are good candidates for quantum computation and communication. Benefit from the development of laser cooling and trapping technology~\cite{raab1987trapping,metcalf2007laser}, neutral atoms can be cooled from room temperature to sub-millikelvin, and subsequently keep atoms trapped at the single atom level for a long time~\cite{miroshnychenko2006quantum}. Now, it is possible to strongly couple single or several trapped atoms to the cavity mode~\cite{neuzner2016interference,welte2017cavity,welte2018photon,kockum2019ultrastrong,chen2019adiabatic,fink2008climbing,fink2009dressed}. Even the ultrastrong coupling has been considered~\cite{gu2017microwave,ribeiro2018polariton,Xie2019generalized}. 

In experiments, detecting whether a single atom is successfully trapped inside the cavity or not is challenging. In general, one can obtain the information of intracavity by detecting the photon leaking from the cavity. Due to the intensity of leakage photons is very weak (mean photon number is less than unity), single-particle detection technology such as the homodyne detection~\cite{horak2003possibility} becomes a powerful tool for single atoms sensing~\cite{bachor2004guide}. To date, there exist several methods to detect single atoms, including direct detection of spontaneously emitted photons~\cite{schlosser2001sub,kuhr2001deterministic}, the observation of the change in the transmission spectrum~\cite{teper2006resonator,nussmann2005vacuum,fortier2007deterministic} and measurement of the polarized photons via the Faraday effect~\cite{terraciano2009photon}. With these current experimental conditions, the quantum statistical properties of photons can also be measured. 

In this paper, we study the interaction of a single atom and the cavity directly driven by a broadband squeezed light. Due to the strong coupling between the atom and cavity, the mean photon number in the cavity, the mean value of $|\langle a a\rangle|$ and the photon number distributions of the photon leaking from the cavity will be significantly changed. Compared with the empty cavity, we show that it is possible to detect single atoms by monitoring the sideway photons, observing the photon number distribution of the one-photon state and measuring the quantum fluctuations of the cavity field. 

\section{System Model}
\begin{figure}[htbp]
	\includegraphics[height=5cm]{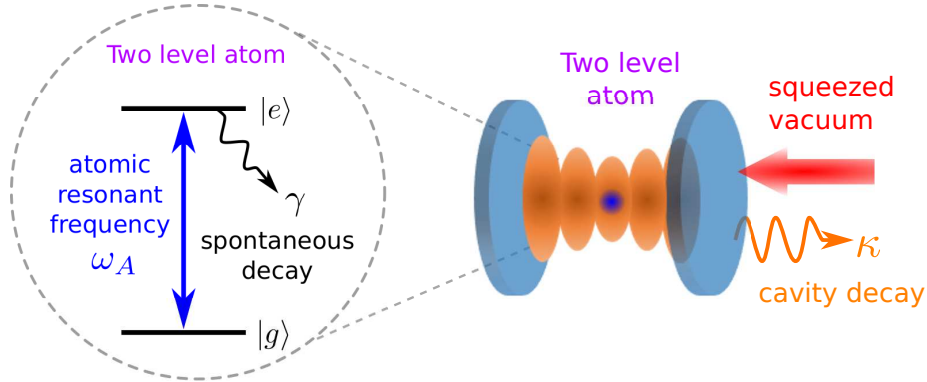}%
	\caption{Sketch of the single atom cavity-QED system driven by a broadband squeezed vacuum with central frequency $\omega_{\rm sq}$. The resonance frequency of this two-level atom is $\omega_{\rm A}=\omega_e-\omega_g$ with $\hbar\omega_\alpha\ (\alpha=e,g)$ being the energy of state $|\alpha\rangle$. Here, $\gamma$ and $\kappa$ are the decay rates of the atom and cavity, respectively.}~\label{fig:fig1}
\end{figure}
In Fig.~\ref{fig:fig1}(a), we first consider a two-level atom trapped in a single-mode cavity with high Q factor, resulting in strong coupling between the cavity mode and atom. The cavity mode frequency is labeled as $\omega_{\rm cav}$ and the atomic resonance frequency is labeled as $\omega_A=\omega_e-\omega_g$ with $\hbar\omega_\alpha\ (\alpha=e,g)$ being the energy of state $|\alpha\rangle$. A broadband squeezed vacuum with central frequency $\omega_{\rm sq}$ is injected into the cavity as shown in panel (a). 

Under the electric dipole approximation, the Hamiltonian of this system can be expressed in the rotating frame of the central frequency of the squeezed field, which reads
\begin{eqnarray}\label{eq:H}
H&=&\Delta_A\sigma_{ee}+\Delta_{\rm C}a^{\dagger}a+g_0(\sigma_{eg}a+\sigma_{ge}a^\dagger),
\end{eqnarray}
where the atomic operator $\sigma_{ij}=|i\rangle\langle j|\ (i,j=\{e,g\})$ and $g_0$ is the coupling strength between the atom and cavity. The annihilation and creation operators of the cavity photons are denoted by $a$ and $a^{\dag}$, respectively. Here, the detunings for the atom and cavity are defined by $\Delta_{\rm A}=\omega_A-\omega_{\rm sq}$ and $\Delta_{\rm C}=\omega_{\rm cav}-\omega_{\rm sq}$, respectively. 
The Eq.~(1) gives only the coherence interaction between the atom and the cavity. The effect of the broadband squeezed radiation is accounted for by using master equation. The terms given by Eq.~(3) below represent the effect of the squeezed drive. 

The evolution of this system is governed by the standard master equation~\cite{agarwal2012quantum}, given by
\begin{equation}\label{eq:master}
\frac{d}{dt}\rho=-\frac{i}{\hbar}[H,\rho]+{\cal L}_{\rm A}(\rho)+{\cal L}_{\rm cav}(\rho),
\end{equation}
where $\rho$ is the density matrix operator. ${\cal L}_{\rm A}(\rho)$ and ${\cal L}_{\rm cav}(\rho)$ are terms describing the decay of the atom and the  cavity driving by a squeezed light, respectively. In general, they have the following forms~\cite{scully1997quantum}: ${\cal L}_{\rm A}(\rho)=\gamma(2\sigma_{eg}^{\dag}\rho \sigma_{eg}-\sigma_{eg}\sigma_{eg}^{\dag}\rho-\rho \sigma_{eg}\sigma_{eg}^{\dag})$ and 
\begin{eqnarray}
{\cal L}_{\rm cav}\rho &=& -\kappa(1+N)(a^\dagger a\rho-2a\rho a^\dagger+\rho a^\dagger a)\nonumber\\
& & -\kappa N(aa^\dagger\rho-2a^\dagger\rho a+\rho aa^\dagger)\nonumber\\
& & +\kappa M (a^\dagger a^\dagger\rho-2a^\dagger\rho a^\dagger+\rho a^\dagger a^\dagger)\nonumber\\
& & +\kappa M^\ast(aa\rho-2a\rho a+\rho aa),
\end{eqnarray} 
with $\gamma$ and $\kappa$ being the damping rates of the atom and cavity, respectively. Here, $N=\sinh^2{(r)}$ is the photon number of the squeezed vacuum with $r$ being the squeezing strength. $M=\cosh{(r)}\sinh{(r)}{\rm e}^{i\phi}$ denotes the two-photon correlation in the injected squeezed vacuum with $\phi$ being the phase of the squeezed vacuum. 
The behavior of atom in a bad cavity ($\kappa\gg\gamma$) driven by a squeezed light has been extensively discussed~\cite{rice1992fluorescent,gardiner1986inhibition}, where the atom-cavity coupling strength $g_0\ll\gamma\ll\kappa$. For good cavities, the effect of the squeezed light in dispersive cavities has been considered~\cite{kono2017nonclassical}, and the reduction of the radiative decay of atomic coherence has been reported~\cite{murch2013reduction}. In this paper, we focus on the strong coupling case and show that the energy exchange between the atom and cavity will result in many appealing phenomena. 
\section{Sensing a single atoms by squeezed drive}
We first discuss one of our key results, i.e., how the squeeze drive can be used to sense a single atom in cavity. To demonstrate this we consider the special case $\Delta_A=\Delta_C=0$. We solve the master equation (2) numerically for a range of the degree of squeezing of the input field. We evaluate the mean photon number $\langle a^\dag a\rangle$ and the photon distribution $P(n)$ in the cavity. In the absence of the atom, these are well known 
\begin{eqnarray}
\langle a^\dag a\rangle=\sinh^2{(r)},\ P_{2n}=\frac{\text{tanh}(r)^{2n}}{\text{cosh}(r)}\frac{{2n}!}{(n!)^22^{2n}},\ P_{2n+1}=0.
\end{eqnarray}
In the presence of the atom, however, the mean photon number and the photon distribution in the cavity change remarkably due to the additional transition pathway induced by the coupling between the atom and cavity. 

\begin{figure}[htbp]
	\includegraphics[height=5cm]{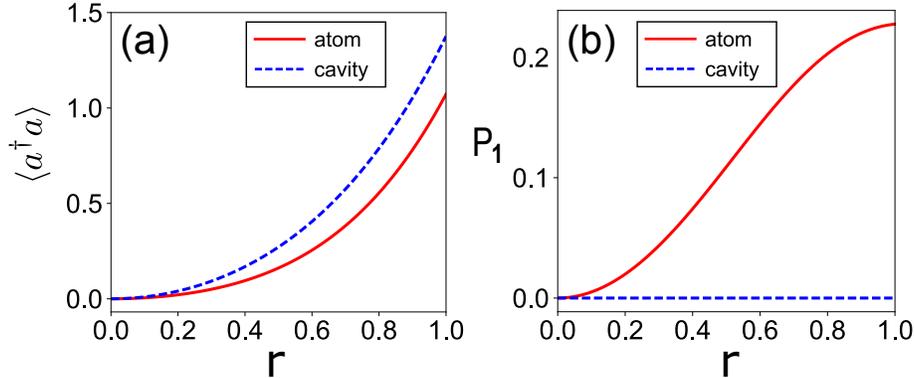}%
	\caption{The mean photon number $\langle a^\dag a\rangle$ [Panel (a)] and $P_1$ [Panel (b)] versus the squeezing parameter $r$. Here, the system parameters are chosen as $\Delta_A=\Delta_C=0$, $g_0/\kappa=15$ and $\gamma/\kappa=1$.
	}
	\label{fig:fig2}
\end{figure}
In Fig.~\ref{fig:fig2} we plot $\langle a^\dag a\rangle$ and $P_1$ as a function of the squeezing parameter. Note that with on atom $P_1=0$. It is clear that the detection of a single photon in the cavity is a sensor for the single atom in the cavity [see Panel (b)]. Note that with increase in the degree of squeezing, the cavity field undergoes significant changes due to the presence of the atom. For small $r$, we can easily understand the process that leads to the excitation of the atom and the presence of a single photon in the cavity, i.e., $|g,0\rangle\underset{radiation}{\overset{Sq}{\rightarrow}}|g,2\rangle\underset{coupling}{\overset{cavity}{\rightarrow}}|e,1\rangle\underset{emission}{\overset{spontaneous}{\rightarrow}}|g,1\rangle$ The applied squeezed radiation produces at least two photons in the cavity. One of these is absorbed by the atom via the Jaynes-Cunnings coupling. The atom can then decay via spontaneous emission. The last transition can be monitored by looking at the radiation on the side of the cavity. 

\begin{figure}[htbp]
	\includegraphics[height=5cm]{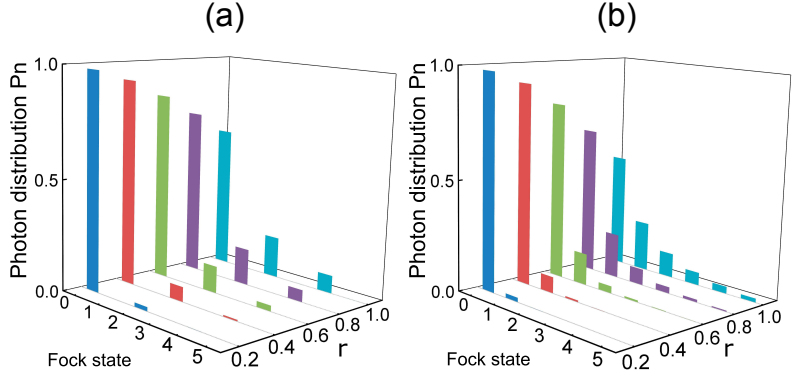}%
	\caption{The photon distribution for arbitrary values of $n$ with different values of $r$. Panels (a) and (b) represent the empty cavity and atom-cavity system, respectively. Here, the system parameters are the same as those used in Fig. 2.}
	\label{fig:fig3}
\end{figure}
In Fig.~(\ref{fig:fig3}) we show the photon distribution for arbitrary values of $n$. The system parameters are the same as those used in Fig. 2. Panel (a) shows that only the states with even number of photons can be detected (i.e., $P_{2k+1}=0$ and $P_{2k}\neq0$) in the absence of the atom~\cite{agarwal2012quantum}. For small $r$, this result can be understood by exploring the excitation pathway $|g,0\rangle\underset{radiation}{\overset{Sq}{\rightarrow}}|g,2\rangle\cdots \underset{radiation}{\overset{Sq}{\rightarrow}}|g,2n\rangle$. With the increase of $r$, more and more states with even number of photons can be detected. Figure (\ref{fig:fig3})b shows how the odd number photon states get excited with the increase in the strength of the squeezing. It is obvious that larger the degree of squeezed light $r$ is, more the states with odd number of photons can be detected. 

\begin{figure}[htbp]
	\includegraphics[height=5cm]{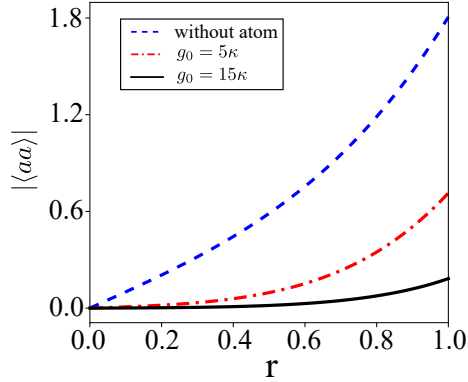}%
	\caption{The phase dependent mean value $|\langle a a\rangle|$ versus the degree of the squeezed light $r$. The blue dashed curves represent the case in the absence of the atom, while other two curves represent the case in the presence of the atom. Here, we choose $g_0/\kappa=5$ for the dash-dotted curve and $g_0/\kappa=15$ for the red solid curve. Other system parameters are the same as those used in Fig. 2.}
	\label{fig:fig4}
\end{figure}
The presence of the atom also changes the phase characteristics of the field in the cavity. This is seen from the plot [Fig.~\ref{fig:fig4}] of the phase dependent mean value $|\langle a a\rangle|$. For the empty cavity, we show that the mean value $|\langle a a\rangle|$ enhances significantly as the value of $r$ increases [see the blue dashed curves in the Fig.~\ref{fig:fig4}]. However, the mean value $|\langle a a\rangle|$ will be suppressed when the atom is in the presence [see other two curves]. The reason for the suppression of $|\langle a a\rangle|$ is the excitation of the atom. As demonstrated by the red solid curve, the suppression of $|\langle a a\rangle|$ can be significantly enhanced if the atom is strongly coupled with the cavity. 

The full quantum statistical characteristics of the cavity field can be described in terms of the Wigner function of the cavity field. The Wigner function is calculated from the solution of the master Eq. (2). 
\begin{figure}[htbp]
	\includegraphics[height=5cm]{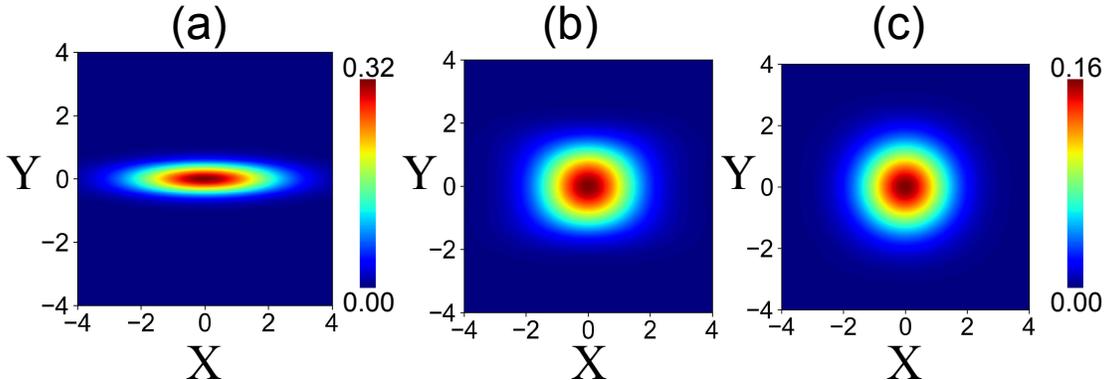}%
	\caption{The Wigner functions of the cavity field. Panel (a) represents the case in the absence of the atom. Panels (b) and (c) represent the cases in the presence of the atom with coupling strengths $g_0/\kappa=5$ and $g_0/\kappa=15$, respectively.}
	\label{fig:fig5}
\end{figure}
In Fig. 5, we plot the Wigner function of the cavity field. In the absence of the atom, the cavity is excited by the two photon process as demonstrated in Fig. 3(a). Thus, the Wigner function is suppressed in one direction [see Fig. 5(a)], implying that the cavity field is squeezed. In the presence of the atom, the excitation of the atom significantly changes the Wigner function of the cavity field as shown in Fig. 5(b) and 5(c). Specifically the presence of a single photon prevent the generation of the squeezed field, resulting in the Wigner function of  a circular aperture under the strong coupling condition [see panel (c)].

An alternate way to understand the interaction of the atom with the squeezed light in the cavity is to make a Bogoliubov transformation on the cavity field. Thus, in the absence of the atom the cavity is in vacuum state of the Bogoliubov operator $b=\cosh(r)a-\sinh{(r)}a^\dag$, $\phi=0$. Then, we can rewrite the master equation (2) as 

\begin{eqnarray}
& &\frac{d\rho}{dt}=-\kappa(b^\dag b \rho-2b\rho b^\dag+\rho b^\dag b)-i[H_I,\rho],\nonumber\\
& &H_I=g\sigma_{eg}(\cosh{(r)}b+\sinh{(r)}b^\dag)+{\rm H.c.}
\end{eqnarray}
In terms of the Bogoliubov modes, the excitation of the atom occurs via the counter rotating terms $\sigma_{eg}b^\dag$. 
\begin{figure}[htbp]
	\includegraphics[height=5cm]{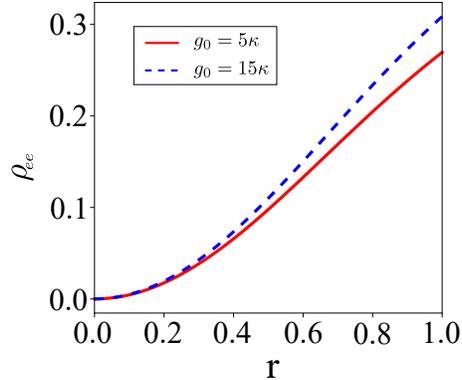}%
	\caption{The population of the atomic excited state $\rho_{ee}$ is plotted as a function of the degree of the squeezed field. Here, we set $\Delta_C=\Delta_A=0$. The blue dashed curve corresponds to the case of $g_0=5\kappa$, but the red solid curve corresponds to the case of $g_0=15\kappa$. Other system parameters are the same as those used in Fig. 2.}
	\label{fig:fig6}
\end{figure}

%

%
Finally, we study the population of the atomic excited state $\rho_{ee}$ by setting $\Delta_C=\Delta_A=0$ and varying the degree of the squeezed field. In Fig. 6, we plot the population of the atomic excited state $\rho_{ee}$ as a function of the degree of the squeezed field. Here, the blue dashed curve corresponds to the case of $g_0=5\kappa$, but the red solid curve corresponds to the case of $g_0=15\kappa$. Other system parameters are the same as those used in Fig. 2. Clearly, the population of the atomic excited state grows quickly (i.e., the atom is excited) with the increase of the degree of the squeezed field. This characteristic implies another way of sensing the atom by detecting the emission of sideway photons once the atom is excited.
\section{conclusion}
In summary, we have studied the single atom-cavity QED system driven by a broadband squeezed light under the strong coupling regime. Due to the energy exchange between the atom and cavity, the mean photon number in the cavity, photon number distributions and quantum fluctuations of the cavity photons are significantly changed. For instance, (I) compared with the empty cavity, the mean photon number in the cavity undergoes a reduction due to the atomic spontaneous emissions, which implies an approach to sensing atoms by detecting the sideway photons emitted by the atom; (II) the mean value of $|\langle a a \rangle|$ can be remarkably suppressed under the strong coupling regime when the atomic resonance frequency is the same as the central frequency of the squeezed light; (III) it is possible to detect photons in the odd photon number states, especially the value of $P_1$ if a single atom is trapped inside the cavity. Based on these interesting properties, we show that the squeezed light is a good candidate for sensing single atoms in the cavity by several potential approaches, such as monitoring the sideway photons, measuring the photon fluctuations and detecting the probability in the one-photon state. These phenomena can be realized with current experimental conditions, and the results presented in this paper may help readers to understand very wide implications of squeezed light interacting with matter.


\section{Funding Information}
The National Key Basic Research Special Foundation (Grant No. 2016YFA0302800); the Shanghai Science and Technology Committee (Grant No. 18JC1410900); the National Nature Science Foundation (Grant No. 11774262);
	

	\bibliography{ref}
	\end{document}